# CHERENKOV LIGHT LATERAL DISTRIBUTION FUNCTION ESTIMATION AS A FUNCTION OF ZENITH ANGLE

**MARWAH M. ABDULSTTAR[1] AND A. A. AL-RUBAIEE[1*]**
[1]Department of Physics, College of Science, Al-Mustansiriyah University, 10052 Baghdad, Iraq.

***Original Research Article***

## ABSTRACT

The simulation of Cherenkov light lateral distribution function (CLLDF) in Extensive Air Showers (EAS) was performed by using the Monte Carlo CORSIKA code for configurations of Tunka-133 EAS Cherenkov array. This simulation was carried out for different primary particles ($e^+$, $e^-$, p, O, Ar and Fe) around the knee region with the energy $3.10^{15}$ eV at different zenith angles. By depending on the Breit-Wigner function a parameterization of CLLDF was reconstructed on the basis of this simulation as a function of the zenith angle. The parameterized CLLDF was verified for four fixed zenith angles in comparison with the simulation that performed using CORSIKA code for each primary particle.

## 1. INTRODUCTION

The Cherenkov radiation effect can be seen when charged particles breaks through dielectric medium, such as air, when exceeds the phase of light in the medium ($v > c/n$), where $c$ is the speed of light; $v$ is the speed of the charged particles and $n$ is the refractive index of the medium. The emission of Cherenkov radiation can be characterized by the superposition of spherical waves using Huygen's principle [1].

As shown in the Fig. 1, the radiation is observed only in a distinctive cone with a specific angle $\theta$ between the particle velocity and the emission direction, where the Cherenkov angle can be obtained as [1]:

$$\cos\theta = \frac{(c/n)t}{(\beta c)t} = \frac{1}{\beta n} \tag{1}$$

Chalenko et al. [2] have used an effective technique of the charged particle background discrimination in the detectors of Cherenkov radiation [3]. While, Vishwanath [4] has extended the field of ultrahigh energy cosmic radiations and gamma ray astronomy, using the atmospheric Cherenkov method, bringing a significant step with the detection of various astronomical sources. On the other hand, E. Korosteleva et al. have proposed a new fitting function for the CLLDF in EAS that obtained from the QUEST data analysis at the EAS-TOP array and from the CORSIKA code simulation [5]. Otherwise, S. Cht. Mavrodiev et al. were proposed a new technique for mass composition and energy spectrum estimation of the primary cosmic ray by depending on the atmospheric Cherenkov light analysis [6]. There are two possible approaches that use Cherenkov photons that created in EAS: first Imaging Air Cherenkov Technique (IACT- measuring angular

*Corresponding author: Email: dr.ahmedrubaiee@gmail.com;*

distribution of Cherenkov light - so called image) [7], where this method working very well in the low energy region, and second by measuring the Cherenkov light density, that is used for energies around the knee region, such as Tunka EAS array.

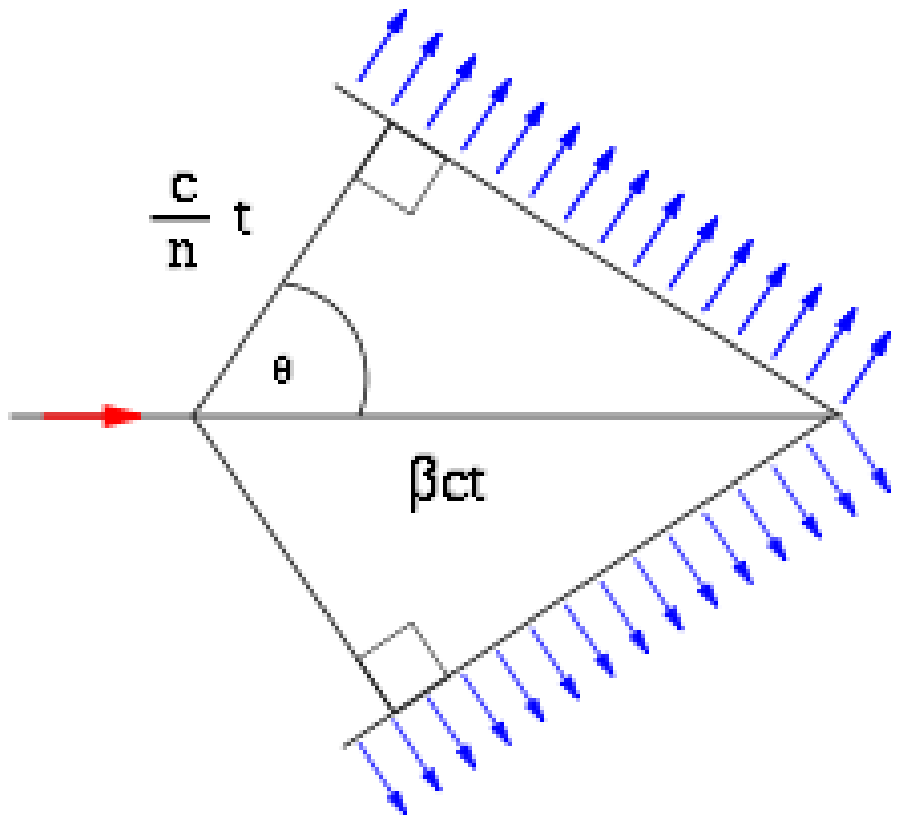

**Fig. 1 Huygen's construction of Cherenkov light [1]**

In the present work it was performed a simulation of CLLDF for the primary particles ($e^+$, $e^-$, p, O, Ar and Fe) by using CORSIKA code for conditions and configurations of the Tunka-133 EAS array around the knee region with the fixed energy $3.10^{15}$ eV at different zenith angles. The parameterization of the numerical simulation results of Cherenkov light density was carried out on the basis of Breit–Wigner function [8], an approach that describing the Cherenkov light lateral function in EAS and analyzing the ability of its application for reconstructing the registered events on the Tunka-133 Cherenkov array. This function was obtained as a function of the zenith angle $\theta$ and verified in comparison with CORSIKA code CLLDF simulation for four fixed zenith angles 5°, 10°, 15° and 20° at fixed primary energy 3 PeV.

## 2. TUNKA-133 EAS ARRAY

Tunka-133 is located in the Tunka valley (50 km from Lake Baikal, Russia) with nearly 1 $km^2$ geometric area which designed to register EAS events from cosmic rays or high energy gamma ray and detect the Cherenkov light of air showers during dark and clear nights [9, 10]. Tunka-133 array consists of 133 wide-angle optical detectors on the basis of PMT EMI 9350 with a hemispherical photocathode of 20 cm diameter. The optical detectors in the array are separated by 19 compact sub arrays that are called clusters, each cluster composed of 7 optical detectors in each one-six hexagonally arranged detectors and one in the center [11]. The distance between the detectors is 85 m and the configuration of the array is shown in Fig. 2 [12].

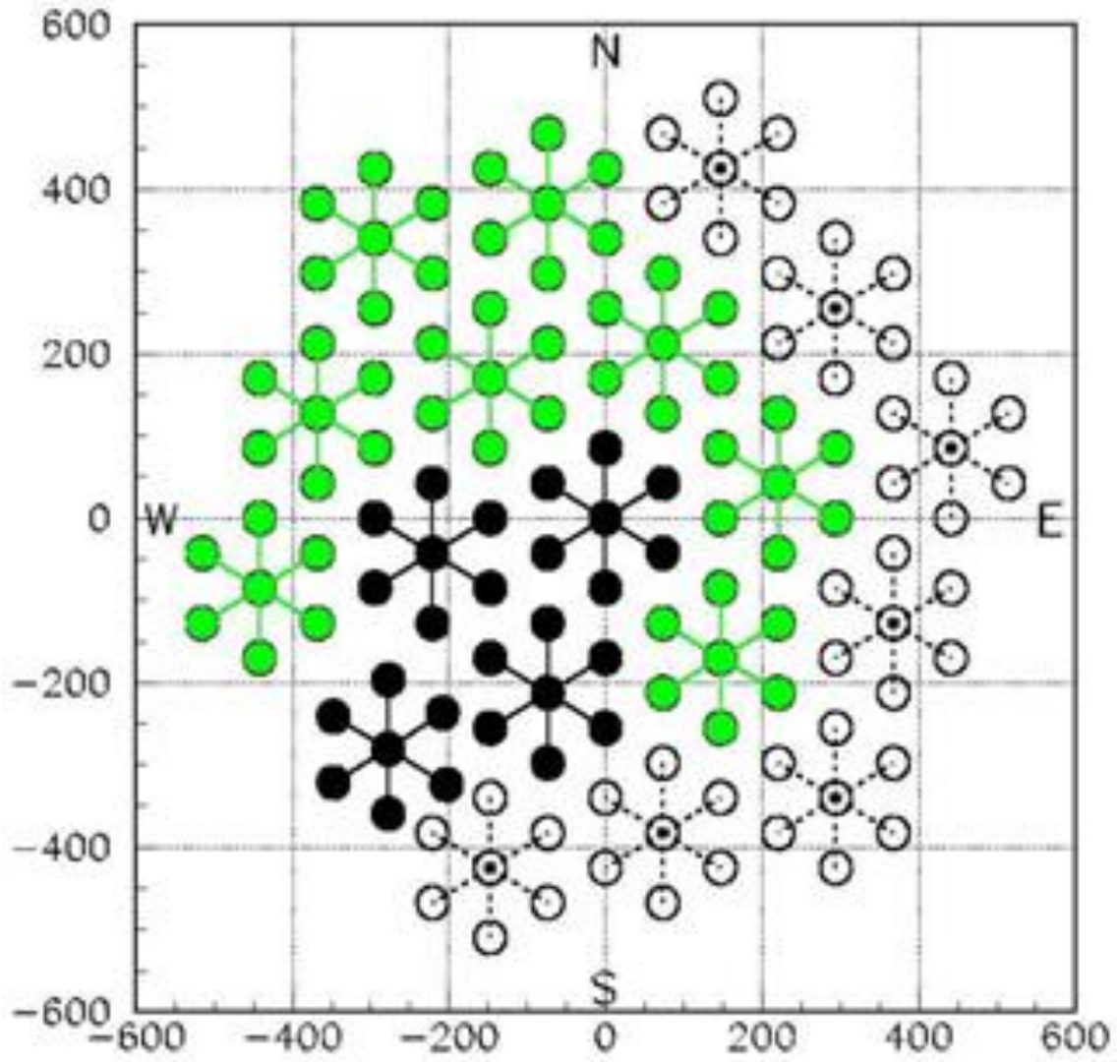

**Fig. 2 Tunka-133 detector configurations [10]**

## 3. THE SIMULATION AND APPROXIMATION MODEL OF CLLDF

### 3.1 The Simulation of CLLDF Using CORSIKA Code

CORSIKA (Cosmic Ray Simulations for KAscade) code is a detailed Monte Carlo program to study the evolution and properties of EAS in the atmosphere [13]. In this work, the simulation of CLLDF was performed by using CORSIKA code with two hadronic models: QGSJET (Quark Gluon String model with JETs) code [14] that was used to model interactions of hadrons with energies greater than 80 GeV and GHEISHA (Gamma Hadron Electron Interaction SHower) code [15] which was used for energies lower than 80 GeV. The simulation of CORSIKA program was performed for the configurations of the Tunka-133 for different primary particles ($e^+$, $e^-$ , p, O, Ar and Fe) with the energy $3.10^{15}$ eV at different zenith angles.

### 3.2 The Parameterization of Simulated CLLDF

The number of Cherenkov photons per the interval of wavelength $(\lambda_1,\ \lambda_2)$ can be obtained from the Tamm and Cherenkov expression [16]:

$$\frac{dN_\gamma}{dx} = 2\pi\alpha \sin^2\theta_r \int_{\lambda_1}^{\lambda_2} \frac{d\lambda}{\lambda^2} =$$

$$= 2\pi\alpha\left(\frac{1}{\lambda_1}-\frac{1}{\lambda_2}\right)\zeta_o\left(1-\frac{E_{th}^2(h)}{E^2}\right)\exp(-h/h_o)\,, \qquad (2)$$

where $\alpha = 1/137$ is the fine structure constant; $\zeta_0 \approx 3.10^{-4}$, $h_0 = 7.5\ Km$; $E_{th} = mc^2\gamma_{th}$, which is the threshold energy of electrons in the atmosphere at the height $h$; $\gamma_{th}$ is the Lorenz factor in the threshold energy that can be given as:

$$\gamma_{th} = \frac{1}{\sqrt{1-\beta^2}} = \frac{1}{\sqrt{1-(1/n(h))^2}} \quad (3)$$

At sea level $n = 1 + \zeta_o$, and the Cherenkov light can be emitted by the electrons that exceed $E_{th}$ i.e. when $\gamma > \gamma_{th} = \frac{E_{th}}{mc^2} \approx 40.8$. By neglecting the absorption of atmospheric Cherenkov radiation, the total number of Cherenkov photons $N_\gamma$, which can be radiated by electrons will be written as [17]:

$$N_\gamma \approx 3.7.10^3 \frac{E_o}{\beta_t}, \quad (4)$$

when

$$N_\gamma \approx 45.10^{10} \frac{E_o}{10^{15} eV}. \quad (5)$$

where $\beta_t$ is the critical energy that equals to the ionization losses at t-unit: $\beta_t = \beta_{ion}\, t_o$ . For electron, $\beta_{ion} = 2.2$ MeV.(g . cm$^{-2}$)$^{-1}$ , $t_o = 37$ g.cm$^{-2}$ and $\beta_t = 81.4$ MeV [18].

Therefore, the number of Cherenkov photons in EAS shower is directly proportional to the primary particle energy. The Cherenkov light density of photons per unit area of the detector to the primary energy of particles is given by the relation [19]:

$$Q_{(E,R)} = \frac{\Delta N_{\gamma\,(E,R)}}{\Delta S}. \quad (6)$$

The relation (6) is utilized for experimental data processing as demonstrated by direct measurements of Cherenkov radiation.

The parameterization of the simulated CLLDF was performed by a function that proposed in Ref. [8] as a function of the distance $R$ from the shower axis and zenith angle $\theta$ that depended on four parameters a, b, σ and $r_o$:

$$Q(\theta,R) = \frac{C\sigma \exp[a-\beta]}{b\left[(R/b)^2 + (R-r_o)^2/b^2 + R\sigma^2/b\right]} \quad (7)$$

Where C=10$^3$ m$^{-1}$ is the normalization constant [20] and $\beta$ is defined as:

$$\beta = R/b + (R-r_o)/b + (R/b)^2 + (R-r_o)^2/b^2 \quad (8)$$

The estimation of Cherenkov light density was performed around the knee region with the energy 3.10$^{15}$ eV at different zenith angles and for different primary particles. Unlike Refs. [21-23], zenith angle dependence of the parameters $a$, $b$, $\sigma$ and $r_o$ permits us to calculate the CLLDF for any zenith angle and fit the LDF which was simulated by CORSIKA code. The zenith angle dependence of the CLLDF parameters was approximated by the relation:

$$K(\theta) = c_0 + c_1 \log_{10}(\theta) + c_2(\log_{10}\theta)^2 + c_3(\log_{10}\theta)^3, \quad (9)$$

here $k(\theta) = a$, log $b$, log σ, log $r_0$; where $\theta$ is the zenith angle; $c_o$, $c_1$, $c_2$ and $c_3$ are coefficients that depended on the zenith angle (see Tables 1 and 2).

**Table 1. Coefficients $c_i$ that determine zenith angle dependence (Eq. 9) of the parameters $a$, $b$, $\sigma$ and $r_o$ for primary positron, electron, and proton for Tunka-133 Cherenkov EAS array**

| $k$ | $c_0$ | $c_1$ | $c_2$ | $c_3$ | $\chi^2$ |
|---|---|---|---|---|---|
| | | | $e^+$ | | |
| $a$ | 9.615.10$^0$ | -4.213.10$^{-2}$ | 3.641.10$^{-2}$ | -8.760.10$^{-3}$ | 2·10$^{-5}$ |
| $b$ | 4.005.10$^{-1}$ | -7.320.10$^{-3}$ | 6.330.10$^{-3}$ | -1.520.10$^{-3}$ | 2·10$^{-5}$ |
| $\sigma$ | -3.088.10$^{-1}$ | -3.312.10$^{-1}$ | 3.112.10$^{-1}$ | -9.386.10$^{-2}$ | 1·10$^{-5}$ |
| $r_o$ | -1.906.10$^{-1}$ | -2.694.10$^0$ | 2.287.10$^0$ | -5.924.10$^{-1}$ | 4·10$^{-4}$ |
| | | | $e^-$ | | |
| $a$ | 9.400.10$^0$ | 6.135.10$^{-1}$ | -6.019.10$^{-1}$ | 1.907.10$^{-1}$ | 5.003·10$^{-6}$ |
| $b$ | 3.897.10$^{-1}$ | 2.474.10$^{-2}$ | -2.366.10$^{-2}$ | 7.390.10$^{-3}$ | 5.001·10$^{-6}$ |
| $\sigma$ | -4.756.10$^{-1}$ | 1.661.10$^{-1}$ | -1.584.10$^{-1}$ | 4.870.10$^{-2}$ | 5.26·10$^{-8}$ |
| $r_o$ | -6.010.10$^0$ | 14.769.10$^0$ | -14.484.10$^0$ | 4.628.10$^0$ | 1.3·10$^{-3}$ |
| | | | p | | |
| $a$ | 8.666.10$^0$ | 2.008.10$^0$ | -1.947.10$^0$ | 5.907.10$^{-1}$ | 4·10$^{-5}$ |
| $b$ | 3.172.10$^{-1}$ | 3.472.10$^{-1}$ | -3.314.10$^{-1}$ | 9.779.10$^{-2}$ | 1·10$^{-3}$ |
| $\sigma$ | 7.962.10$^{-1}$ | -3.940.10$^0$ | 3.846.10$^0$ | -1.180.10$^0$ | 28·10$^{-5}$ |
| $r_o$ | 4.203.10$^{-1}$ | -5.359.10$^0$ | 5.203.10$^0$ | -1.559.10$^0$ | 1·10$^{-3}$ |

**Table 2. Coefficients $c_i$ that determine zenith angle dependence (Eq. 9) of the parameters *a, b, σ* and $r_o$ for primary oxygen, Argon, and iron nuclei for Tunka-133 Cherenkov EAS array**

| | | O | | | |
|---|---|---|---|---|---|
| *k* | $c_0$ | $c_1$ | $c_2$ | $c_3$ | $\chi^2$ |
| *a* | $9.118.10^{0}$ | $2.580.10^{-1}$ | $-2.628.10^{-1}$ | $8.707.10^{-2}$ | $4.53\cdot10^{-6}$ |
| *b* | $3.783.10^{-1}$ | $1.113.10^{-1}$ | $-1.077.10^{-1}$ | $3.364.10^{-2}$ | $5.11\cdot10^{-6}$ |
| *σ* | $-3.261.10^{-1}$ | $-3.573.10^{-2}$ | $3.676.10^{-2}$ | $-1.212.10^{-2}$ | $8.75\cdot10^{-8}$ |
| $r_o$ | $-3.786.10^{0}$ | $9.099.10^{0}$ | $-9.462.10^{0}$ | $3.191.10^{0}$ | $1\cdot10^{-3}$ |
| | | Ar | | | |
| *a* | $8.767.10^{0}$ | $2.097.10^{-1}$ | $7.964.10^{-1}$ | $-6.005.10^{-1}$ | $6.1\cdot10^{-3}$ |
| *b* | $3.935.10^{-1}$ | $1.167.10^{-2}$ | $-9.240.10^{-3}$ | $2.220.10^{-3}$ | $8.21\cdot10^{-7}$ |
| *σ* | $-3.601.10^{-1}$ | $-1.804.10^{-1}$ | $1.716.10^{-1}$ | $-5.252.10^{-2}$ | $8.04\cdot10^{-7}$ |
| $r_o$ | $1.210.10^{1}$ | $-4.106.10^{1}$ | $4.073.10^{1}$ | $-12.930.10^{0}$ | $4\cdot10^{-3}$ |
| | | Fe | | | |
| *a* | $1.519.10^{1}$ | $-1.946.10^{1}$ | $20.483.10^{0}$ | $-7.020.10^{0}$ | $4.3\cdot10^{-3}$ |
| *b* | $8.283.10^{-1}$ | $-1.320.10^{0}$ | $1.360.10^{0}$ | $-4.550.10^{-1}$ | $1.3\cdot10^{-3}$ |
| *σ* | $4.347.10^{-1}$ | $-2.694.10^{0}$ | $2.777.10^{0}$ | $-9.285.10^{-1}$ | $1\cdot10^{-4}$ |
| $r_o$ | $-2.512.10^{0}$ | $5.601.10^{0}$ | $-6.529.10^{0}$ | $2.397.10^{0}$ | $4\cdot10^{-3}$ |

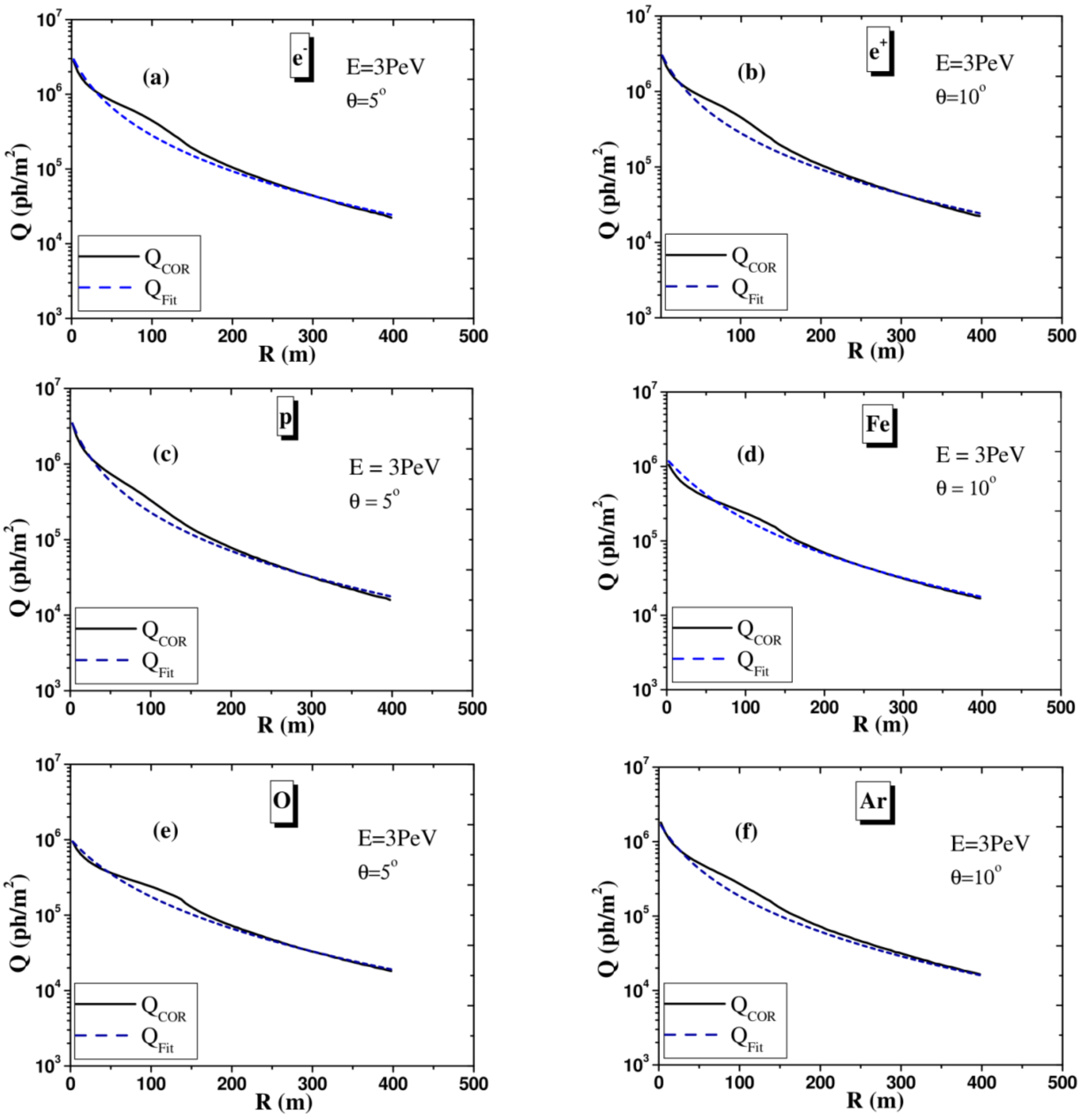

**Fig. 3 Comparison of the simulated CLLDF using CORSIKA code for Tunka-133 array conditions at the energy 3 PeV at θ=5° and 10° (solid lines) and one approximated using Eqs. (7), (8) and (9) (Dash lines) for: (a) $e^-$; (b) $e^+$; (c) p; (d) Fe; (e) O and (f) Ar**

## 4. RESULTS AND DISCUSSION

Fig. 3 demonstrates the results of the simulated CLLDF (solid curves) in comparison with that parameterized (Dash lines) using Equations (7) and (8) with their parameters (Equation (9)) for different primary particles ($e^-$, $e^+$, p, Fe, O and Ar) at zenith angles θ=5° and 10°. While in Fig. 4 was shown the results of the simulated CLLDF (solid curves) in comparison with that parameterized at zenith angles θ=15° and 20° (Dash lines).

The parameterized CLLDF in Figs. 3 and 4 for slant EAS showers slightly differs from the simulated CLLDF that depended on the Tunka133 EAS array configuration. This difference was identified by minimizing the function:

$$\Delta = \sum_i \left[ \frac{Q_{fit}(\theta, R_i)}{Q_{COR}(R_i)} - 1 \right]^2 \rightarrow \min \quad , \qquad (10)$$

where the results for each primary particle are illustrated in Table 3 for two fixed zenith angles.

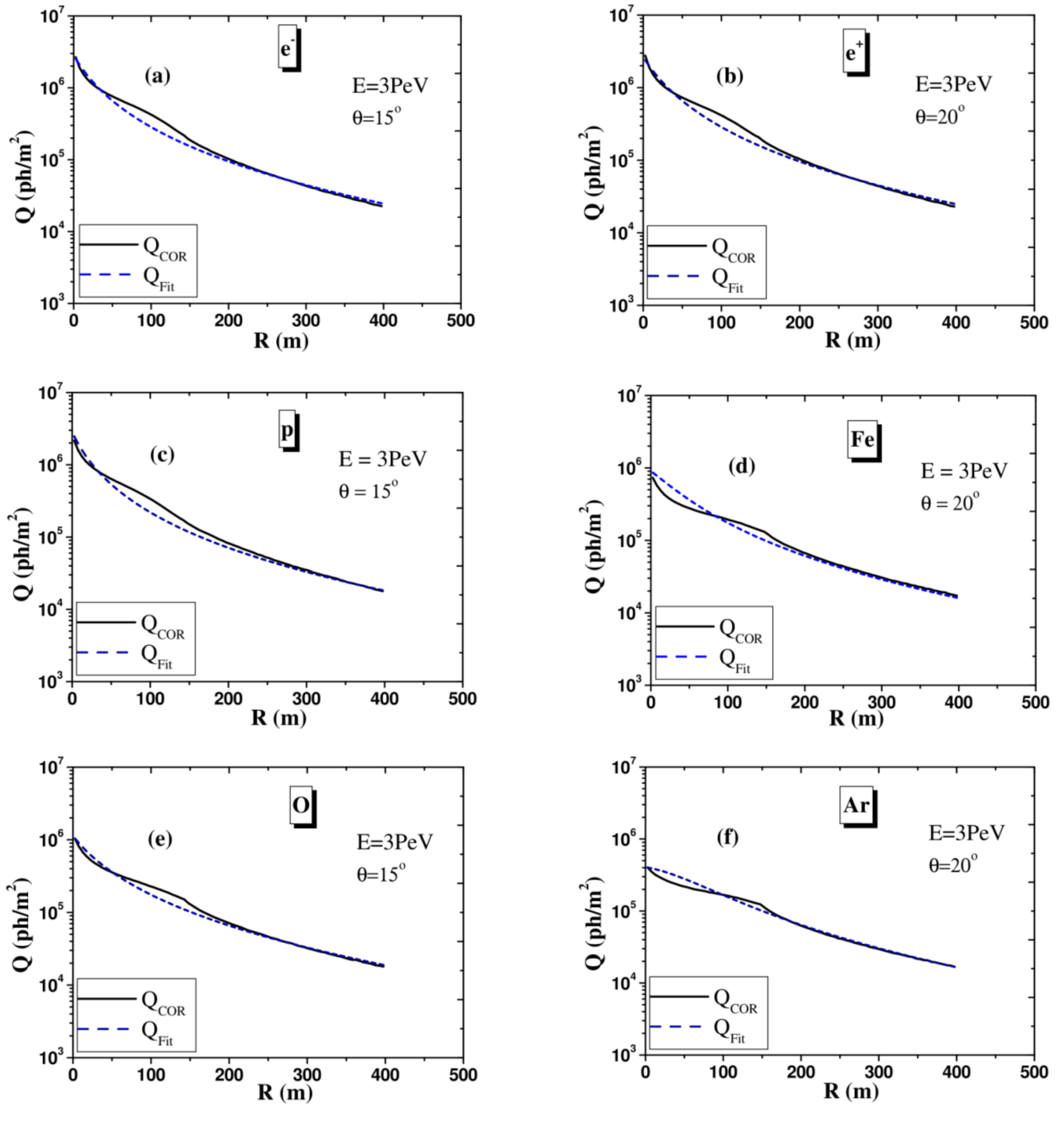

**Fig. 4 Comparison of the simulated CLLDF with CORSIKA code for Tunka-133 array configurations with the energy 3 PeV at θ=15° and 20° (solid lines) and one approximated using Eqs. (7), (8) and (9) (Dash lines) for: (a) $e^-$; (b) $e^+$; (c) p; (d) Fe; (e) O and (f) Ar**

**Table 3. The results of $\Delta_{min}$ that calculated using Eq. (10) for each primary particle at at the primary energy 3 PeV and zenith angles 5° and 10°**

| | $\Delta_{min}$ | |
|---|---|---|
| **Particle type** | $\theta=5^o$ | $\theta=10°$ |
| $e^+$ | $1.308.10^{-2}$ | $3.363.10^{-2}$ |
| $e^-$ | $3.007.10^{-3}$ | $6.151.10^{-3}$ |
| p | $1.095.10^{-2}$ | $5.282.10^{-3}$ |
| O | $2.322.10^{-3}$ | $2.322.10^{-3}$ |
| Ar | $1.366.10^{-2}$ | $1.208.10^{-4}$ |
| Fe | $2.090.10^{-3}$ | $4.109.10^{-3}$ |

## 5. CONCLUSION

The Cherenkov light lateral distribution function in EAS was simulated using CORSIKA code for conditions and configurations of the Tunka-133 Cherenkov array for different primary particles ($e^+$, $e^-$, p, O, Ar and Fe) around the knee region with the energy $3.10^{15}$ eV at different zenith angles. The lateral distribution function of Cherenkov radiation, which was proposed in Ref. [6] was developed by approximating the results of the numerical simulation of CLLDF as a function of the distance from the shower axis and zenith angle $\theta$. The parameterized CLLDF by using Equations (7), (8) and (9) for each primary particle was verified with four different zenith angles (5°, 10°, 15° and 20°). This verification was gave a good agreement in comparison with the simulated CLLDF in the core distance range 2.5-40 m and from 150-400 m. There is a partial disagreement in the range 50-200 m from the shower axis because of the fluctuation of particle densities in the CORSIKA code simulation especially. The main advantage of the given approach consists of the possibility to make a library of LDF samples which could be utilized for analysis of real events which detected with the EAS array and reconstruction of the primary cosmic rays energy spectrum and mass composition.

### AUTHOR'S CONTRIBUTION

This work was carried out in collaboration between all authors. Author SEA designed the study, wrote the protocol and interpreted the data. Author BA anchored the field study, gathered the initial data and performed preliminary data analysis. Authors OA and LEI managed the literature searches and produced the initial draft. All authors read and approved the final manuscript.

## COMPETING INTERESTS

Authors have declared that no competing interests exist.